# Measuring Financial Sentiment to Predict Financial Instability: A New Approach based on Text Analysis

Rickard Nyman, Paul Ormerod[*] and David Tuckett
Centre for the Study of Decision Making Uncertainty, University College, London

February 2015

- **Corresponding author p.ormerod@ucl.ac.uk**; University College, London WC1E 6BT, UK

**Abstract**

Following the financial crisis of the late 2000s, policy makers have shown considerable interest in monitoring financial stability. Several central banks now publish indices of financial stress, which are essentially based upon market related data.

In this paper, we examine the potential for improving the indices by deriving information about emotion shifts in the economy. We report on a new approach, based on the content analysis of very large text databases, and termed directed algorithmic text analysis (DATA). It draws on a social-psychological theory of decision-making under uncertainty to focus on just two classes of emotion in narratives – those prompting either excitement about gain or anxiety about loss. The algorithm identifies, very rapidly, shifts through time in the relations between two these core emotional groups. The method is robust. The same word-list is used to identify the two emotion groups across different studies. Membership of the words in the lists has been validated in psychological experiments. The words consist of everyday English words with no specific economic meaning.

Initial results show promise. An emotion index capturing shifts between the two emotion groups in texts potentially referring to the whole US economy improves the one-quarter ahead consensus forecasts for real GDP growth. More specifically, the same indices are shown to Granger cause both the Cleveland and St Louis Indices of Financial Stress.

1. **Introduction**

Following the financial crisis of the late 2000s, policy makers have shown considerable interest in monitoring financial stability. Several central banks now publish indices of financial stress. For example, in early 2010, the St. Louis Fed created the STLFSI, which uses 18 weekly data series to measure financial stress in the market. The STLFSI is constructed from seven interest rate series, six yield spreads and five other indicators. Each of these variables captures some aspect of financial stress. The average value of the index, which begins in late 1993, is designed to be zero. Thus, zero is viewed as representing normal financial market conditions. Values below zero suggest below-average financial market stress, while values above zero suggest above-average financial market stress.

The Cleveland Fed has developed the CFSI. This index takes components that quantify individual aspects of the system and combines them into a single value. Specifically, the CFSI is constructed using daily data from 11 components reflecting four financial sectors: credit markets, equity markets, foreign exchange markets, and interbank markets. The overall financial system is complex and comprises many individual markets of varying size and significance. Stress in any of these four could carry over to others, affecting the system at large. Detailed comparisons of the various indices are given by Kliesen et al. (2012) and Manamperi (2013).

The purpose of this paper is to examine the potential for improving the indices by incorporating a new measure of emotional shifts. The idea that emotion is a important driver of the economy has a long history in economics, certainly as far back as Keynes, who coined the phrase 'animal spirits' in his *General Theory* (1936).

A number of recent papers attempt to analyse news and other texts from digital sources and then to explore the effects on the economy. For example, Romer and Romer (2010) analysed text sources (including presidential speeches) to measure the likely effects of tax change announcements on subsequent economic behaviour. Dominguez and Shapiro (2013) analysed newspaper and media sources to identify events prompting narrative shifts (for instance developments in the Euro crisis) to explore if they could account for the slowness of the economic recovery. Baker et al (2013, 2014) constructed an uncertainty index, counting the number of uncertainty words that appeared in the content of news media. Soo (2013) quantified the positive and negative tone of housing news in local newspaper articles. She used those measures to test the role of sentiment in the run-up and crash of housing prices that instigated the great financial crisis of 2008.

A feature of the first three papers is that the underlying hypothetical model is that "economic" news (information likely to influence an agent with rational expectations) is postulated as the causal factor, whereas in Soo's study it is emotion expressed in the context of housing news. Our interest is in how far emotional shifts might exert an influence on the economy independently of "economic" events reported in the news.

In this paper, we report some promising results obtained using a new approach, based on the content analysis of very large text databases, and termed directed algorithmic text analysis (DATA). It draws on a social-psychological theory of decision-making under uncertainty to focus specifically on the emotion in narratives. To do this the algorithm identifies, very rapidly, shifts through time in the relations between just two classes of emotion in the narratives it searches – those prompting either approach (excitement about gain) or avoidance (anxiety about loss).

Using the method, we construct a time series for the US economy over the period from 1996 to 2014, measuring the net balance of the two emotion groups in narratives, obtained by analysing the text of Reuters news feeds sourced in New York and Washington. For purpose of description, we term this series 'relative sentiment shift' or RSS for short.

Section 2 sets out a relatively brief description, with supporting references, of the Methodology. This section is necessary because of the innovative nature of the approach. Section 3 provides a combined Results and Discussion, and section 4 gives a short Conclusion.

## 2. Methodology

### 2.1 Decision making under uncertainty

The methodology which we describe is a way of making operational a theory of human decision making under uncertainty. That is to say, decision making in situations in which the probability distribution of outcomes is itself either unknown or inherently uncertain.

The dominant paradigm of expectation formation within economics is that of rational expectations. This requires considerable knowledge on the part of agents of the 'true' model which describes the operation of the economy. Agents either are already in possession of the relevant model, or discover it through some form of Bayesian learning. However, in many situations, especially in macroeconomics, there is unresolved uncertainty about the model itself. For example, looking back to the policy debates in the immediate aftermath of the collapse of Lehman Brothers in September 2008, prominent economists, including Nobel Laureates, could be found on both sides of the argument as to whether or not to allow banks and other financial institutions to fail. It is hard to imagine that these groups of protagonists had the same model of the economy in mind.

More generally, within the statistics literature, there is a widespread understanding that model uncertainty is often an inherent feature of reality. It may simply not be possible to decide on the 'true' model. Chatfield (1995) is a widely cited paper on this topic. In an economic context, Onatski and Williams (2003), for example, in a survey for the European Central Bank of sources of uncertainty, concluded that "The most damaging source of uncertainty for a policy maker is found to be the pure model uncertainty, that is the uncertainty associated with the specification of the reference model". Gilboa et al. (2008) note that "the standard expected utility model, along with Bayesian extensions of that model, restricts attention to beliefs modeled by a single probability measure, even in cases where no rational way exists to derive such well-defined beliefs".

In short, in situations in which there is uncertainty about the true model which describes the system, it may not possible for agents to form rational expectations. As a result, agents are uncertain about the probability distribution of potential outcomes. The psychological theory which motivates the direction of our search for the emotions of 'excitement' and 'anxiety'(see section 2.3 below), was developed to describe how agents gain the confidence to take decisions in such circumstances, rather than simply being paralysed into inactivity in the face of uncertainty (Tuckett and Nikolic, 2015).

**2.2 The text data base**

The Thomson-Reuters News archive (1996-2014) consisted (at the time of our analysis) of over 17 million English news articles. Reuters provide extensive documentation of the various columns in each file. We make use of the 'date', 'language', 'text', 'attribution' and 'tags' fields. The 'date' field contains the publication date, the 'language' field the publication language, the 'text' field the main article text, the 'attribution' tag states the publisher and the 'tags' field contains a comma separated list of article tags provided by Reuters.

We consider only articles with Reuters as the attribution and English as the language. Where the text field starts with 'NEW YORK' or 'WASHINGTON', we consider the articles as US focused. Where the text feed starts with 'LONDON', we consider the articles as UK focused. and exclude them from this particular set of analyses. The US data is analysed in terms of its relationship to the real economy in Nyman et al. (2014). In all instances, to avoid articles tagged as 'Sport', 'Weather' or 'Human interest' we remove articles with tags SPO (sports), ODD (human interest) or WEA (weather) within the 'tags' field.

Articles are available daily, and this is the time interval at which we search the data. Results from the searches can be readily aggregated over time to generate monthly or quarterly data.

A final point to note is that the Thomson-Reuters news feed is meant to be precisely that. Journalists are meant to report the news and not write opinion pieces. The emotional content of the data base compared to, for example, a set of brokers' circulars, is therefore low.

**2.3 Conviction narrative theory and searching the data base**
    **2.3.1   The context**

Our approach extends the literature cited in the section 1 in three ways.

First, as mentioned, the choice of which emotions to measure is guided by psychological theory. The existing literature is essentially atheoretical with respect to the emotion terms which are searched for. This is not to say that the terms are implausible, but they are quite general and can also be seen as *ad hoc*. Our approach is based upon a specific social-psychological theory of agent decision making under uncertainty, so that it is directed theoretically.

Second, as mentioned, we are careful to separate the possible role of emotion from the role of economic news. The specific set of words we search for were chosen solely for their role as words which evoke the two emotion groups the theory specified. They are not words which have an economic meaning outside their everyday human meaning. We then used standard experimental techniques in psychology to validate the emotional content of the words. This inherent element of direction to the search process leads us to describe the technique as Directed Algorithmic Text Analysis (DATA).

The third and final point is simply that the searches are carried out, in contrast to those in several of the papers cited above, using modern computer science technology. This improves the speed of search by many orders of magnitude compared, for example, to the human based searched which feature in much of the previous literature in economics. For instance, the entire Reuters news data base, containing millions of articles, can be searched on a standard personal computer in a matter of a few minutes.

### 2.3.2 Conviction narrative theory

The psychological theory of action under uncertainty, conviction narrative theory, is described in detail elsewhere Tuckett and Nikolic (2015), Chong and Tuckett (2014), Tuckett et al (2014), Tuckett (2014). Essentially, it focuses on two emotional groups. One group of emotions is based around anxiety. This acts to inhibit action. The other group around excitement acts as an amplifier to action. Anxiety causes people to react to uncertainty by abandoning or deferring commitment. So when economic outcomes become difficult to anticipate in a secure way, economic action will be potentially paralysed. However, distrust and potential paralysis can be overcome through the development of conviction about an uncertain action's likely success. Excitement about the outcome can increase agents' commitment. The point is that when trying to decide what to do human actors make use of the evolved capacities of affect and narrative simulation to impose structure on their predictions by developing embodied simulations of possible futures. Conviction narratives make sense of past and present, project potential futures and support the sense of their accuracy going forward. To act to approach an investment project agents require conviction, they need to envisage outcomes that attract (generating excitement) and also repel worries that things will not work out (anxiety).

### 2.3.3 Word lists

Uncertainty creates the possibility of success and failure and goes on doing so. Faced with it, agents will tend to imagine scenarios that create a mixture of excitement and anxiety. The specific words we chose to indicate the two theoretically important emotion groups were first selected by expert judgment as representative and then subjected to an experimental validity test in which random samples of words from the two lists were presented to financially-literate participants. Asked to rate whether the words they were shown expressed anxiety about loss or excitement about gain, or neither, their answers – despite the fact they were given only a general context – strongly agreed with the expert classification (Strauss 2013).

A crucial point to emphasise again is that these word lists do **not** contain any specific economic terms such as 'crisis' or 'boom'. They are what we might term regular English words, words in everyday use in a wide variety of contexts which convey the emotions of either 'excitement' or 'anxiety', validated as such in psychological experiments. We can think of them as being orthogonal to the economic data which we analyse. Examples of the words are given in Nyman et al. (op. cit.)

What we call relative sentiment shift (RSS) is a summary statistic of the two emotional traits extracted from text data by counting the two types of words. For the summary statistic of a collection of texts T we count the number of occurrences of excitement words and anxiety words and then scale these numbers by the total text size in number of articles. To arrive at a single statistic, based on the underlying theory of conviction narratives, we subtract the anxiety statistic from the excitement statistic.

$$Sentiment[T] = \frac{|Excitement| - |Anxiety|}{size[T]}$$

We make the obvious remark that an increase in this relative emotion score is due to an increase in excitement and/or a decrease in anxiety.

It is also possible to focus on a specific concept, such as 'liquidity', by way of example. The search will then only analyse articles which contain the word 'liquidity'. A number of different metrics can be used to measure proximity of any of the words in the excitement and anxiety lists to the word 'liquidity'. For example, the net balance of the emotional words can be counted in any particular article which contains the word 'liquidity'. Alternatively, the words can be counted only if they are in the same sentence as 'liquidity', or within a specified number of characters of 'liquidity'

Given the size of the data base being analysed, it is not necessary to control for possible negations of these words, e.g. 'not anxious'. However, we did examine explicitly whether the results are influenced if 'not' is present. There are standard ways in algorithmic text search to detect the proximity of words. We found that the theoretical prior, namely that negations make no difference, is not rejected by the evidence. Given that this paper is essentially about economics and not computer science, details of this can be obtained from the authors.

The simplicity of this method is intentional. It makes very clear what we measure and how we measure it and allows us to bring a range of statistical techniques to bear on correlations.

## 3. Results and Discussion

### 3.1 Initial overview

Figure 1 below plots the RSS series obtained from the analysis of the Reuters news feed data base described above. The series, to recap, measures the difference between the number of excitement words and the number of anxiety words, divided by the total number of articles. The data is available on a daily basis, which in Figure 1 is aggregated temporally onto a monthly basis from January 1996 through June 2014. The series is then normalized with mean zero and standard deviation of one.

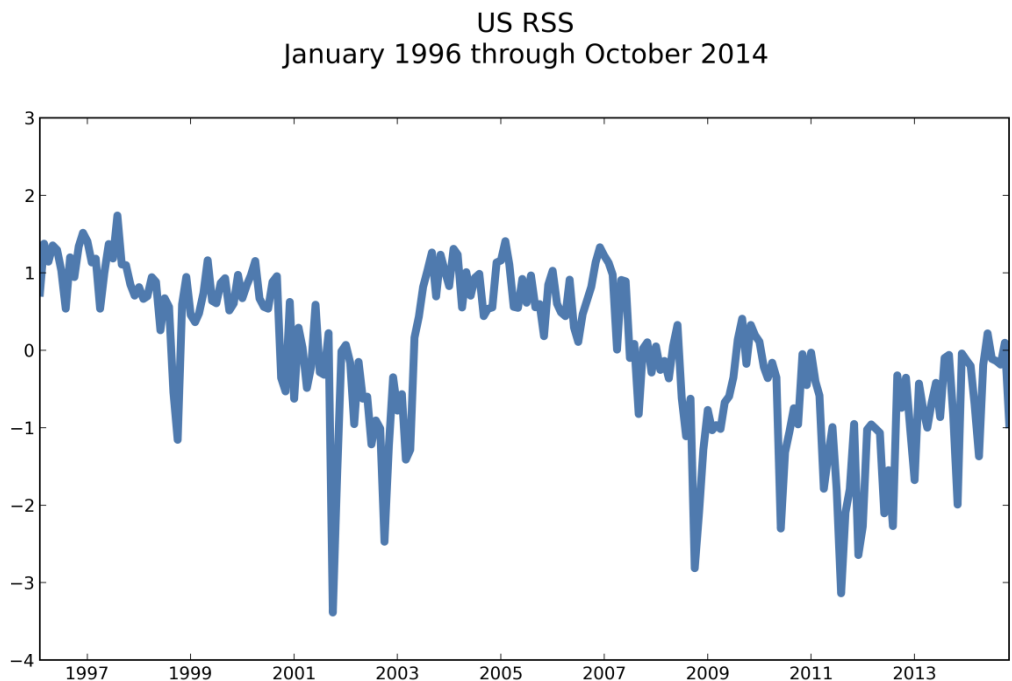

**Figure 1**   *Overall relative sentiment shift ('animal spirits') in the US. The series is the difference between the number of 'excitement' and the number of 'anxiety' words in the Thomson-Reuters news feed, divided by the total number of articles, and then normalised*

A detailed analysis of the ability of the RSS series to improve understanding of the real economy in the US is provided in Nyman et al. (op.cit.). It should be noted in particular that it Granger causes series such as the Economic Policy Uncertainty Index of Baker et al. (op.cit.) and quarterly US GDP growth, with no evidence of causation in the reverse direction.

Here, we demonstrate an illustration of initial credibility of the RSS series by showing its potential to improve the one-quarter ahead forecasting record of real GDP growth in the United States.

The Survey of Professional Forecasters is the oldest quarterly survey of macroeconomic forecasts in the United States. The survey began in 1968 and was conducted by the American Statistical Association and the National Bureau of Economic Research. The Federal Reserve Bank of Philadelphia took over the survey in 1990. Data on the consensus forecast for one-quarter ahead real GDP growth is available at

http://www.philadelphiafed.org/research-and-data/real-time-center/survey-of-professional-forecasters/. A discussion of the historical accuracy of the forecasts, for both GDP and other economic variables, is given in Stark (2010).

The consensus forecasts over time are unbiased. However, they are able to account for only a relatively small fraction of the overall variance in quarterly real GDP growth. We confirm this finding in the literature by regressing quarterly real GDP growth in quarter *t* on the consensus forecast for quarter *t* made in quarter *t-1* over the period 1996Q2 through 2014Q3.

**Table 1**: Regression of the actual values of quarterly growth in real US GDP (DLGDP) on the consensus forecast made in the previous quarter (SPF)

|                     | Dependent variable: |
|---------------------|---------------------|
|                     | DLGDP               |
| SPF                 | 1.123*** (0.292)    |
| Constant            | -0.430 (0.804)      |
| Observations        | 74                  |
| R2                  | 0.170               |
| Adjusted R2         | 0.159               |
| Residual Std. Error | 2.469 (df = 72)     |
| F Statistic         | 14.770*** (df = 1; 72) |

Note: *p<0.1; **p<0.05; ***p<0.01

We now add to equation (1) our RSS measure in quarter *t-1*.

**Table 2**: Regression of the actual values of quarterly growth in real US GDP (DLGDP) on the consensus forecast made in the previous quarter (SPF) and the relative sentiment shift series (RSS)

|  | Dependent variable: |
| --- | --- |
|  | DLGDP |
| SPF | 0.829*** |
|  | (0.304) |
| RSS | 0.765** |
|  | (0.299) |
| Constant | 0.325 |
|  | (0.829) |
| Observations | 74 |
| R2 | 0.240 |
| Adjusted R2 | 0.219 |
| Residual Std. Error | 2.379 (df = 71) |
| F Statistic | 11.231*** (df = 2; 71) |
| Note: | *p<0.1; **p<0.05; ***p<0.01 |

The RSS variable is statistically significant from zero, and the adjusted R squared increases from 0.159 in the equation without RSS to 0.219 when it is included, and incremental effect of 38 per cent.

The same results hold when the third vintage estimate of GDP is used rather than the latest data used in equations (1) and (2), and the results are set out in Tables 1a and 2a in the Appendix.

### 3.2 Granger causality of RSS, the Cleveland and St Louis Financial Stress Indices

We report results of tests of Granger causality between the overall RSS series described above and both the CFSI and the STLFSI using monthly data over the period January 1996 through September 2014.

We use the methodology described in Toda and Yamamoto (1996). In outline, in investigating Granger causality between any two series, this is as follows:

1. Check the order of integration of the two series using Augmented Dickey-Fuller (Said and Dickey 1984; p-values are interpolated from Table 4.2, p. 103 of Banerjee et al. 1993) and the Kwiatowski-Phillips-Schmidt-Shin (1992) tests. Let $m$ be the maximum order of integration found.

2. Specify the VAR model *using the data in levelled* form whatever was found in step 1 determine the number of lags to use with standard method. We use the Akaike Information Criteria

3. Check the stability of the VAR (we use OLS-CUSUM plots, which are reported in the Appendix).

4. Test for autocorrelation of residuals. If autocorrelation is found, increase the number of lags until it goes away. We use the multivariate Portmanteau- and Breusch-Godfrey tests for serially correlated errors. Let p be the number of lags then used

5. Add m extra lags of each variable to the VAR

6. Perform Wald tests with null being that the first p lags of the independent variable have coefficients equal to 0. If this is rejected, we have evidence of Granger-causality from the independent to dependent variable.

We used the statistical program R to carry out the analysis, and the various packages used to carry out the above Toda-Yamamoto procedure are documented in the Appendix.

**Table 3:** Augmented Dickey-Fuller and Kwiatowski-Phillips-Schmidt-Shin tests for stationarity

| Variable | ADF | p-value | KPSS | p-value |
|---|---|---|---|---|
| RSS Level | -2.91 | 0.195 | 2.32 | < 0.01 |
| RSS Diff | -6.95 | < 0.01 | 0.025 | > 0.1 |
| CFSI Level | -2.95 | 0.175 | 0.649 | 0.018 |
| CFSI Diff | -8.24 | < 0.01 | 0.044 | > 0.1 |
| STLFSI Level | -2.81 | 0.236 | 0.623 | 0.02 |
| STLFSI Diff | -5.57 | < 0.01 | 0.049 | > 0.1 |

Note: *p<0.1; **p<0.05; ***p<0.01

Note: the lag order for the ADF tests is 6 and the truncation lag parameter for the KPSS tests is 3

The null hypothesis in the ADF test is that there is a unit root, so that we look to reject the null for the variable to be stationary, and the null in the KPSS is stationarity, so we look to not reject the null. The results of Manamperi (op. cit.) that the variables are non-stationary in level form are confirmed by Table 3. All the variables are I(1).

In terms of deciding the lag length in the VAR models, we examine the pairs of variables (RSS, CFSI) and (RSS, STLFSI). In each case we include a constant but no trend, and test for lags from 1 through 20. Full details of the AIC values are in Table 3a in the Appendix. For (RSS, CFSI) the AIC is minimised at 5 lags and for (RSS, STLFSI) at 4 lags.

After selecting the appropriate number of lags, we check parameter stability using empirical fluctuation process plots (which we denote as OLS-CUSUM, following Ploberger and Kramer (1992). The charts are reported in the Appendix, and there are no problems with stability.

The next step is to test the residuals of the two VAR models for autocorrelation. In neither case is the null hypothesis of no autocorrelation rejected.

**Table 4**: Chi-square tests of the null hypothesis of no autocorrelation of the residuals in the VAR models

| VAR model | Portmanteau | d.f. | Breusch-Godfrey | d.f. |
|---|---|---|---|---|
| CFSI/RSS | 46.17 | 44 | 23.83 | 20 |
| STLFSI/RSS | 43.36 | 48 | 13.02 | 20 |

Finally, we report the Wald tests for Granger causality between the RSS series and each of the Financial Stress Indices.

**Table 5**: Wald test of Granger causality between Financial Stress Indices and the general Relative Sentiment Shift Index

```
Direction      Chi-Sq d.f.   p-value

CFSI -> RSS     9.7    5     0.086*
RSS -> CFSI    16.3    5     0.0061***
STLFSI -> RSS   3.4    4      > 0.5
RSS -> STLFSI  24.5    4     6e-05***
```

Note: *p<0.1; **p<0.05; ***p<0.01

There is weak evidence of Granger causality from the CFSI to the RSS, but the p-value in the test for causality from RSS to CFSI is an order of magnitude less. There is no evidence of causality from the STLFSI to the RSS, and very strong evidence of causality from RSS to the STLFSI

## 3.3 Granger causality of Liquidity RSS, the Cleveland and St Louis Financial Stress Indices

We illustrate further the potential of the approach by constructing an RSS measure based on the word 'liquidity'. We again analyse the Thomson-Reuters news feed, this time only using articles in which the word 'liquidity' appears. We use as a metric 100 characters either side of the word 'liquidity'. This means that the words in the excitement and anxiety lists are only counted if they appear within 100 characters either side of the word 'liquidity'.

Using this metric, the data is rather thin during the first few years in which the data is available to us. We therefore carry out this analysis starting in January 1999, again ending in September 2014.

A full set of results, as reported in section 3.2, is available on request from the authors. Here, we simply report the relevant final table of section 3.2, namely the Wald tests of Granger causality.

**Table 6**: Wald test of Granger causality between Financial Stress Indices and the Relative Sentiment Shift Index focused on 'liquidity'

| Direction | Chi-Sq | d.f. | p-value |
|---|---|---|---|
| CFSI -> RSSLIQ | 0.6 | 2 | 0.76 |
| RSSLIQ -> CFSI | 5.7 | 2 | 0.058** |
| STLFSI -> RSSLIQ | 0.7 | 2 | 0.71 |
| RSSLIQ -> STLFSI | 11.5 | 2 | 0.0032*** |

Note: $^*p<0.1$; $^{**}p<0.05$; $^{***}p<0.01$

The results using the RSS measure focused on liquidity are qualitatively very similar to those of Table 5. However, there is no causality from the Cleveland index to the RSS measure defined in this way.

The results in Table 6 illustrate how the general RSS measure, based on a count of all relevant emotion words in the Thomson-Reuters news feed, can be modified and become more focused on concepts which maybe more directly relevant to the specific issue being examined. 'Liquidity', for example, may be a more focused way of looking at the issue of financial stress rather than the more general RSS measure. But clearly, there is more to do in this area.

## 4. Conclusion

The results here are illustrative and are not meant to be definitive. In particular, we do not attempt to develop an index which incorporates the emotion data.

However, the results suggest quite clearly that there is potential to improve current asset-price based indices of financial stability by incorporating time series information on emotion in the economy. We show that emotion series have forward looking power in understanding the evolution over time of two of the existing indices of financial stability. Both the general Relative Sentiment Shift series, and one which is based on the word 'liquidity', Granger-cause both the Cleveland and St Louis Indices of Financial Stress.

It is important to emphasise once again that the words which we use to measure the emotions of excitement and anxiety are words which are in everyday use in English, and do not contain examples of words which have more specific economic meaning, or which are used frequently in economic commentaries.

Useful results have been obtained when emotions are measured in a large text data base which is deliberately meant to simply report news and not offer opinions or emotions. The exploration of other data bases, such as brokers' reports, in which emotional content is intended, may offer scope to strengthen the findings.

# Appendix

The packages in R used in the Toda-Yamamoto procedure to investigate Granger causality are as follows:

- *tseries* – we use the two functions *adf.test* and *kpss.test* (the Augmented Dickey-Fuller test and Kwiatkowski-Phillips-Schmidt-Shin test respectively) to check if series are stationary or contain unit roots
- *vars* – we use the function *VARselect* to compute the Akaike Information Criteria for VAR(p) processes with p from 1 through 20. We use the *VAR* function for estimating a VAR(p) process. We use the function *serial.test* to compute the multivariate Portmanteau- and Breusch-Godfrey tests for serially correlated errors in a VAR(p) process. We use the function *stability* to compute empirical fluctuation processes according to the OLS-CUSUM method
- *aod* – we use the function *wald.test* to perform the Wald tests for Granger causality

We use the *lm* function available in the *stats* package to fit linear models.

## Supplementary Tables

**Table 1a**: Regression of the 3$^{rd}$ vintage estimate of values of quarterly growth in real US GDP (DLGDP3EST) on the consensus forecast made in the previous quarter (SPF)

|  | Dependent variable: |
|---|---|
|  | DLGDP3EST |
| SPF | 1.066*** |
|  | (0.264) |
| Constant | -0.061 |
|  | (0.726) |
| Observations | 73 |
| R2 | 0.186 |
| Adjusted R2 | 0.175 |
| Residual Std. Error | 2.230 (df = 71) |
| F Statistic | 16.262*** (df = 1; 71) |
| Note: | *p<0.1; **p<0.05; ***p<0.01 |

**Table 2a**: Regression of the 3$^{rd}$ vintage estimnate of quarterly growth in real US GDP (DLGDP3EST) on the consensus forecast made in the previous quarter (SPF) and the relative sentiment shift series (RSS)

|  | Dependent variable: |
|---|---|
|  | DLGDP3EST |
| SPF      | 0.760*** |
|          | (0.271)   |
| RSS      | 0.793*** |
|          | (0.266)   |
| Constant | 0.722     |
|          | (0.737)   |
| Observations | 73 |
| R2 | 0.278 |
| Adjusted R2 | 0.258 |
| Residual Std. Error | 2.115 (df = 70) |
| F Statistic | 13.485*** (df = 2; 70) |

Note: *p<0.1; **p<0.05; ***p<0.01

Note that the RSS data are not subject to revision.

**Table 3a**: Values of AIC criterion for selection of lags in the VAR models containing (CFSI, RSS) and (STLSFI, RSS)

| Lags | CFSI | STLFSI |
|---|---|---|
| 1 | -2.667 | -3.129 |
| 2 | -2.694 | -3.283 |
| 3 | -2.748 | -3.306 |
| 4 | -2.740 | -3.339 |
| 5 | -2.756 | -3.327 |
| 6 | -2.755 | -3.296 |
| 7 | -2.742 | -3.275 |
| 8 | -2.745 | -3.245 |
| 9 | -2.721 | -3.215 |
| 10 | -2.698 | -3.193 |
| 11 | -2.682 | -3.173 |
| 12 | -2.679 | -3.156 |
| 13 | -2.673 | -3.165 |
| 14 | -2.643 | -3.152 |
| 15 | -2.619 | -3.130 |
| 16 | -2.615 | -3.104 |
| 17 | -2.592 | -3.092 |
| 18 | -2.572 | -3.069 |
| 19 | -2.577 | -3.062 |
| 20 | -2.541 | -3.052 |

**Supplementary Figures**

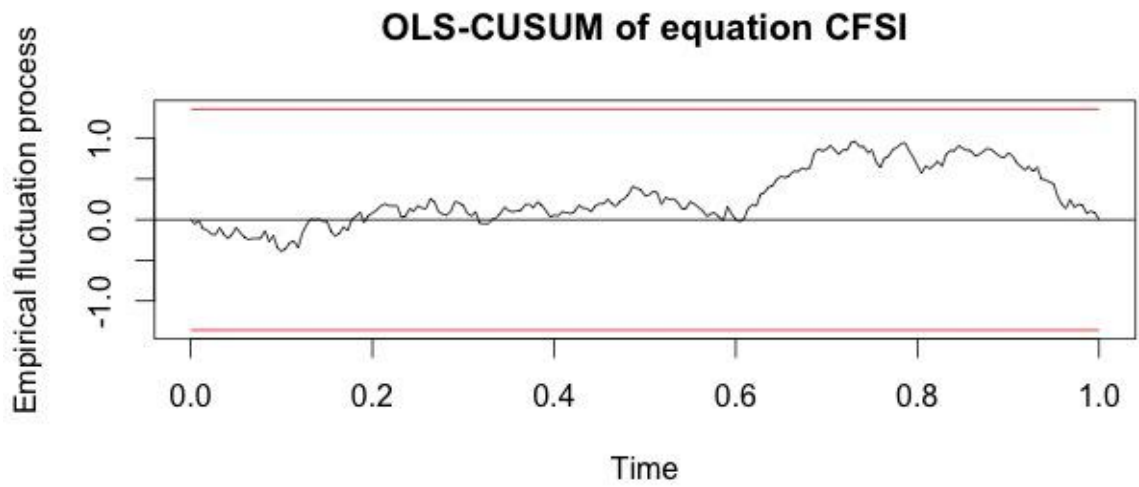

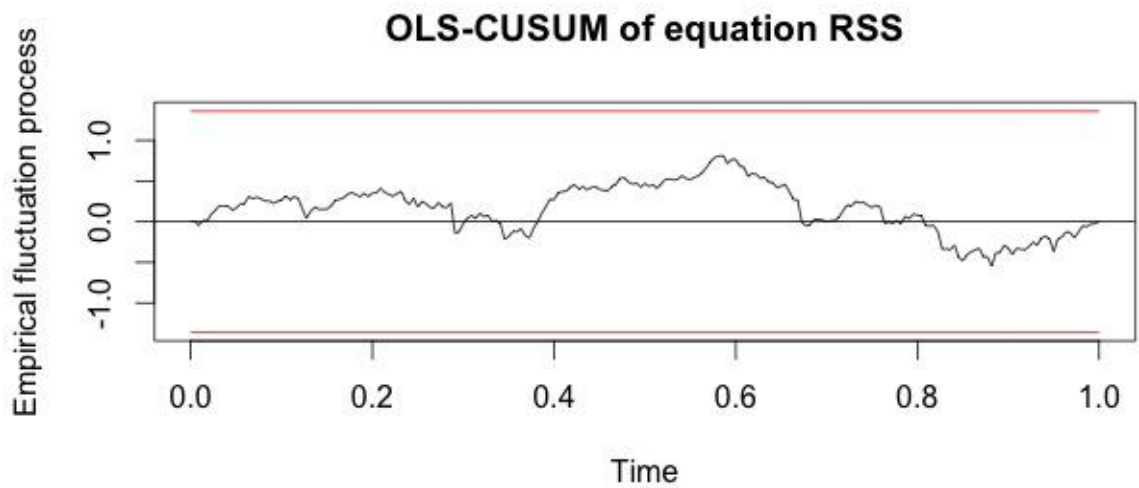

**Figure 1a:** OLS-CUSUM plots from the VAR model containing CFSI and RSS

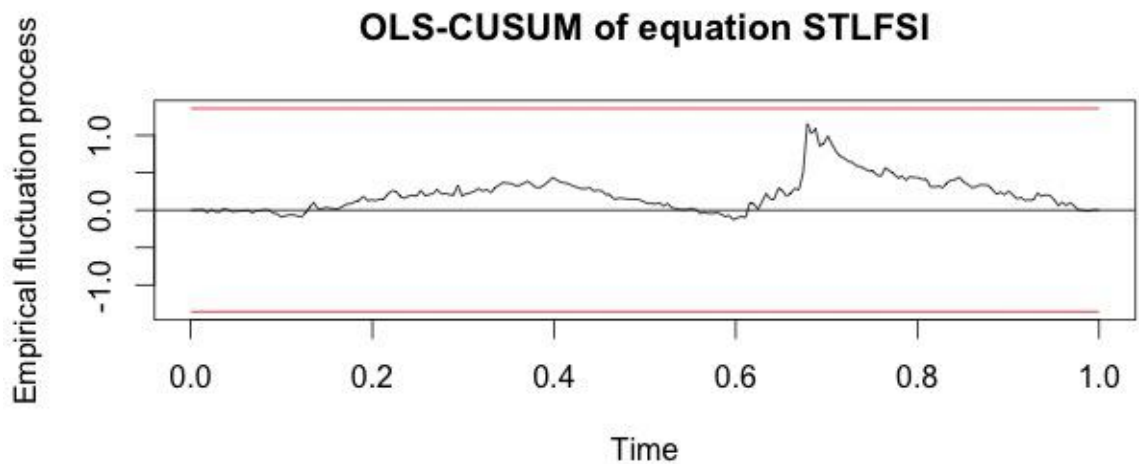

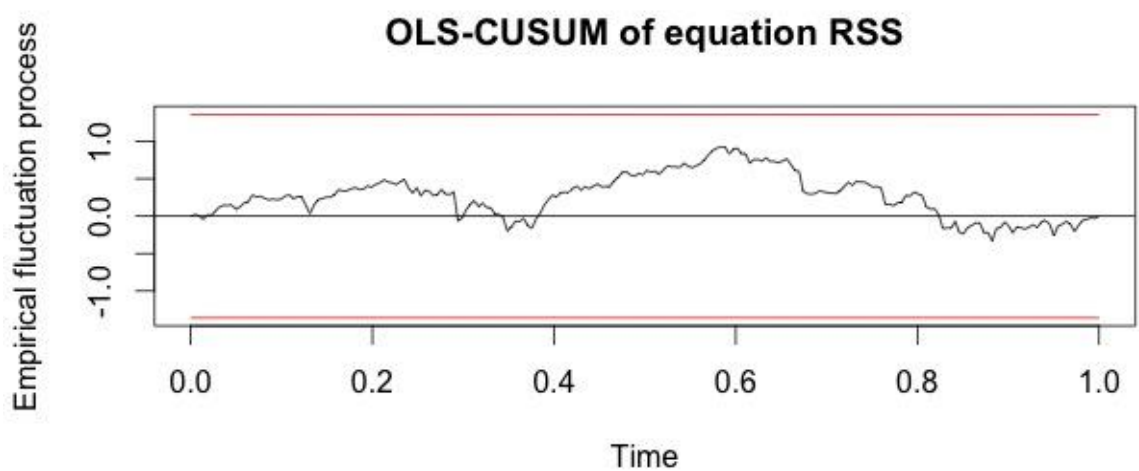

**Figure 1b:** OLS-CUSUM plots from the VAR model containing STLFSI and RSS